\documentclass{emulateapj}
\usepackage{subfigure}
\usepackage{graphicx}

\def\be{\begin{equation}}
\def\ee{\end{equation}}
\def\ba{\begin{eqnarray}}
\def\ea{\end{eqnarray}}
\newcommand{\degree}{\ensuremath{^\circ}}

\shorttitle{Radio Pulsar Intermittency}
\shortauthors{Deneva et al.}

\begin{document}

\title{Arecibo Pulsar Survey Using ALFA: Probing Radio Pulsar Intermittency and Transients}

\author{
J.~S.~Deneva$^{1,*}$, 
J.~M.~Cordes$^{1}$, 
M.~A.~McLaughlin$^{2}$,
D.~J.~Nice$^{3}$, 
D.~R.~Lorimer$^{2}$,
F.~Crawford$^{4}$,
N.~D.~R.~Bhat$^{5}$, 
F.~Camilo$^{6}$, 
D.~J.~Champion$^{7}$, 
P.~C.~C.~Freire$^{8,2}$, 
S.~Edel$^{2}$,
V.~I.~Kondratiev$^{2}$,
J.~W.~T.~Hessels$^{9,10}$, 
F.~A.~Jenet$^{11}$, 
L.~Kasian$^{12}$, 
V.~M.~Kaspi$^{13}$, 
M.~Kramer$^{14,15}$,
P.~Lazarus$^{13}$,
S.~M.~Ransom$^{16}$, 
I.~H.~Stairs$^{13,8,5}$,
B.~W.~Stappers$^{14}$,
J.~van~Leeuwen$^{10,11}$, 
A.~Brazier$^{1}$,
A.~Venkataraman$^{9}$,
J.~A.~Zollweg$^{17}$,
and S.~Bogdanov$^{13}$
\\ \vspace{10pt} \today}

\affil{$^{1}$Astronomy Dept., Cornell Univ., Ithaca, NY 14853} 
\affil{$^{2}$Department of Physics, West Virginia Univ.,
Morgantown, WV 26506}
\affil{$^{3}$Physics Dept., Bryn Mawr College, Bryn Mawr, PA
19010}
\affil{$^{4}$Department of Physics and Astronomy, Franklin and Marshall College, P.O. Box 3003, Lancaster, PA 17604-3003}
\affil{$^{5}$Centre for Astrophysics and Supercomputing, Swinburne
Univ.~of Technology, Hawthorn, Victoria 3122, Australia} 

\affil{$^{6}$Columbia Astrophysics Laboratory, Columbia Univ., New York, NY 10027}
\affil{$^{7}$ATNF-CSIRO, Epping, NSW 1710, Australia} 
\affil{$^{8}$NAIC, Arecibo Observatory, PR 00612}
\affil{$^{9}$Netherlands Institute for Radio Astronomy (ASTRON), Postbus 2, 7990 AA Dwingeloo, The Netherlands}
\affil{$^{10}$Astronomical Institute ``Anton Pannekoek,'' Univ.
of Amsterdam, 1098 SJ Amsterdam, The Netherlands} 
\affil{$^{11}$Center for Gravitational Wave Astronomy, Univ. of Texas at Brownsville, TX 78520} 

\affil{$^{12}$Dept.~of Physics and Astronomy, Univ.~of British
Columbia, Vancouver, BC V6T 1Z1, Canada} 

\affil{$^{13}$Dept.~of Physics, McGill Univ., Montreal, QC H3A
2T8, Canada}

\affil{$^{14}$Univ. of Manchester, Jodrell Bank Centre for Astrophysics, 
Alan Turing Building, Manchester M13 9PL, UK} 

\affil{$^{15}$Max-Planck-Institut f\"{u}r Radioastronomie, Auf dem Huegel 69, 53121 Bonn, Germany}

\affil{$^{16}$NRAO, Charlottesville, VA 22903}  
\affil{$^{17}$Center for Advanced Computing, Cornell Univ., Ithaca, NY 14853}  
\affil{$^\ast$E-mail: deneva@astro.cornell.edu}


\begin{abstract}
We present radio transient search algorithms, results, 
and statistics from the ongoing Arecibo Pulsar ALFA (PALFA) 
survey of the Galactic plane. We have discovered seven objects through a search for isolated dispersed pulses.  All of these objects are Galactic
and have measured periods between 0.4 and 4.7~s. 
One of the new discoveries has a duty cycle 
of 0.01\%, smaller than that of any other radio pulsar. We discuss the impact of selection 
effects on the detectability and classification of intermittent sources, 
and compare the efficiencies of periodicity and 
single-pulse searches for various pulsar classes. 
For some cases we find that the apparent intermittency is likely to
be caused by 
off-axis detection or
a short time window that selects only a few bright pulses and favors detection
with our single-pulse algorithm.  In other cases, the intermittency appears
to be intrinsic to the source.  No transients were
found with dispersion measures large enough to require that
they originate from sources outside our Galaxy. 
Accounting for the on-axis gain of the ALFA system, as well as the low gain but large solid-angle coverage of far-out sidelobes, we use the results of the survey so far to place limits on the amplitudes and event rates of transients of arbitary origin.

\end{abstract}

\section{Introduction}

Radio pulsars show a wide variety of modulations of their pulse amplitudes, including bursts and nulls, that affect their detectability in surveys. 
Phenomena seen in some pulsars include short-period nulling, in which a 
pulsar is not detected for several pulse periods, only to reappear with 
full strength (\citealt{Backer70}); eclipses, in which a companion star 
or its wind or magnetosphere absorbs or disperses the pulsar signal (\citealt{Fruchter88},\citealt{Stappers96},\citealt{Lyne93},\citealt{Kaspi04}); 
long-term nulling or intermittent behavior, in which a pulsar is 
quiescent for days or weeks (\citealt{Kramer06}); and rotating radio transients (RRATs, \citealt{McLaughlin06}), pulsar-like objects 
from which only occasional radio bursts are detected. This paper 
describes analysis of a large-scale survey using the Arecibo telescope
that is sensitive to both periodic and aperiodic signals.

RRATs were first discovered in archive Parkes Multibeam survey data (\citealt{McLaughlin06}). 
Eleven objects, with periods ranging from 0.7 to 7~s and pulse widths of $2 - 30$~ms, were found using a single pulse search algorithm (\citealt{McLaughlinRev07}). The longer periods of RRATs compared with the general pulsar population suggest similarities with the X-ray populations of X-ray dim isolated neutron stars (XDINSs) and magnetars. RRAT J1819$-$1458 has been detected at X-ray energies (\citealt{McLaughlin07}) with properties that are similar to those of XDINSs and high magnetic field radio pulsars. 


A different type of pulse modulation is observed in the case of pulsars emitting giant pulses. Such pulses are tens to thousands of times brighter and an order of magnitude or more narrower than the average pulse (see \citealt{Knight06} for an overview). Giant pulses from the Crab pulsar have substructure on timescales of 2~ns (\citealt{Hankins03}), and PSR~B1937+21 emits giant pulses as narrow as 16~ns (\citealt{Popov04}). Giant micro-pulses from the Vela pulsar have widths $\sim 50 \mu$s (\citealt{Johnston01}), and the slowly rotating pulsars PSR~B1112+50, PSR~B0031$-$07, and PSR~J1752+2359 occasionally emit bright pulses which are $1 - 10$~ms wide, $5 - 30$ times narrower than the average pulse (\citealt{Knight06}). 
The detection of giant pulses is a potentially powerful method for finding 
extragalactic pulsars too distant for their normal emission to 
be detectable by periodicity searches (\citealt{McLaughlin03}). 

A variety of energetic phenomena other than pulsar emission can give rise to fast transients potentially detectable in radio pulsar surveys. Within the Solar System, transient radio events may be generated by energetic particles impacting the Earth's atmosphere, solar flares, and decameter radio flares originating in Jupiter's atmosphere. Analogously to the latter, extrasolar planets with strong magnetic fields are expected to be detectable in the 10-1000~MHz range (\citealt{Farrell99}, \citealt{Lazio04}, \citealt{Zarka01}). Magnetic activity on the surfaces of brown dwarfs and particle acceleration in the magnetic fields of flare stars are also known radio flare progenitors (\citealt{Berger01}, \citealt{Berger02}, \citealt{Garcia03}, \citealt{Jackson89}). Gamma-ray bursts are predicted to have detectable radio emission at $\sim 100$~MHz (\citealt{Usov00}, \citealt{Sagiv02}), and radio flares have been observed from some X-ray binaries (\citealt{Waltman95}, \citealt{Fender97}). Among the most energetic and exotic events in the Universe, supernovae, merging neutron stars and coalescing black holes may produce wide-band radio bursts detectable at extragalactic distances (\citealt{Hansen01}).

In this paper we describe an ongoing survey for pulsars and transient radio sources with the Arecibo telescope. The survey addresses outstanding questions about the nature and emission mechanisms of intermittent radio sources. In \S~\ref{section_surveyparams} we present the PALFA survey parameters, and in \S~\ref{section_spmethods} we describe the single pulse search and radio frequency interference excision algorithms which are part of the survey data processing pipeline. Section~\ref{section_palfastats} contrasts PALFA detection statistics on known pulsars and new discoveries, and \S~\ref{section_biases} examines selection effects influencing the detection and classification of transient sources. In \S~\ref{section_intermittency}, we apply an intermittency measure method for comparing the efficiency of periodicity and single pulse pulsar searches. In \S~\ref{section_objects}, we discuss the properties of individual intermittent objects discovered by PALFA. In \S~\ref{section_faroutmodel} and \S~\ref{section_constraints}, we apply constraints derived from the survey sensitivity and results to the detectability of various energetic phenomena expected to emit radio bursts.  Finally, in \S~\ref{section_conclusions}, we present our main conclusions.

\section{PALFA Survey Observations}\label{section_surveyparams}

\subsection{Survey Parameters}

The PALFA survey started in 2004, shortly after the installation of the seven-beam ALFA receiver on the Arecibo telescope. The survey searches for pulsars and transients in the inner and outer Galactic plane regions accessible to Arecibo (see below). 

The ALFA receiver is well-suited for survey observations, allowing simultaneous data collection from seven fields, each $\sim 3.5'$ (FWHM) across. Taking into account the hexagonal arrangement of the beams on the sky and the near sidelobes, the combined power-pattern is approximately $24' \times 26'$ (\citealt{palfa1}). We observe a 100 MHz passband centered on 1440~MHz in each of the seven telescope beams. Wideband Arecibo Pulsar Processors (WAPPs; \citealt{Dowd00}) are used to 
synthesize 256-channel filterbanks
spanning these bands at intervals of 64~$\mu$s. During observations, full-resolution data are recorded to disk and, in parallel, decimated down to a time resolution of 1~ms and searched for periodic signals and single pulses by a quick-look processing pipeline running in real time at the Arecibo Observatory (\citealt{palfa1}). This approach allows for immediate discovery of relatively bright pulsars with periods longer than a few milliseconds. Searching full-resolution data allows detection of millisecond pulsars and narrower single pulses and is done offline at participating PALFA institutions as the processing is much more computationally intensive. 

Table~\ref{pointing_stats} lists various ALFA system and survey parameters, including the sky area corresponding to processed and inspected data reported on in the present paper and the total sky area observed to date (see \citealt{palfa1} for a detailed explanation of other parameters). Standard observation times are 268~s for inner Galaxy pointings ($30\degree \lesssim l \lesssim 78\degree, |b| \leq 5\degree$)  and 134~s for outer Galaxy pointings ($162\degree \lesssim l \lesssim 214\degree, |b| \leq 5\degree$). Some early observations had a duration of 134~s and 67~s for inner and outer Galaxy pointings, respectively. The quoted system temperature of 30~K is measured looking out of the Galactic plane. The initial threshold of $5\sigma$ for the single pulse search is used when selecting events from dedispersed time series based on their signal-to-noise ratios only, before any filtering is applied. While there are a significant number of events due to random noise above this threshold, identification of an event as a genuine pulse takes into account not only its signal-to-noise ratio but also the fact that it is detected at a contiguous range of trial dispersion measures, which is in general not true of spurious events. Thus weak pulses can be correctly identified, while they would be excluded if the threshold was set according to Gaussian noise statistics.

\begin{deluxetable}{lccc}
\tablecolumns{4}
\tablewidth{0pc}
\tablecaption{PALFA survey parameters.\label{pointing_stats}}
\tablehead{\colhead{Parameter} & \colhead{Value}}
\startdata
Center frequency (GHz) & 1.440 \\
Total bandwidth (MHz) & 100 \\
Channel Bandwidth (MHz) & 0.39 \\
Sampling time ($\mu$s) & 64 \\
Nominal system temperature (K) & 30 \\ 
Gain (K/Jy) & \\
\phn\phn Center pixel & 10.4 \\
\phn\phn Ring pixels & 8.2 \\
Beam width (') & \\
\phn\phn Main beam (1 pixel) & 3.5 \\
\phn\phn Main beam plus near sidelobes (1 pixel) & 18 \\
\phn\phn Main beam plus near sidelobes (7 pixels) & $24 \times 26$ \\
Inner Galaxy & \\
\phn\phn Observation time per pointing (s) & 268, 134 \\
\phn\phn Observed (deg$^2$) & 156 \\ 
\phn\phn Processed\tablenotemark{a} (deg$^2$) & 99 \\ 
Anticenter & \\
\phn\phn Observation time per pointing (s) & 134, 67 \\
\phn\phn Observed (deg$^2$) & 119 \\ 
\phn\phn Processed\tablenotemark{a} (deg$^2$) & 87 \\ 
Detection threshold ($\sigma$) & \\
\phn\phn Single pulse search (initial threshold) & 5 \\
\phn\phn Fourier Transform periodicity search & 7.5
\enddata
\tablenotetext{a}{Full-resolution data processed with the Cornell pulsar search code.}
\end{deluxetable}

\subsection{Survey Sensitivity}

Here we compute the maximum distance at which sources of a given luminosity are be detectable by the PALFA survey, $D_{\rm max}$. We then compare the sensitivity of the PALFA survey to previous work done with the Parkes Multibeam system (\citealt{PMBsurv}, \citealt{McLaughlin06}).

The rms noise in a radio transient search, where the effective observation time is equal to the pulse width, is
\be
\sigma_n = \frac{S_{\rm sys}}{\sqrt{N_{\rm pol}~\Delta f~W}},
\ee
where $S_{\rm sys}$ is the system-equivalent flux density, $N_{\rm pol} = 2$ is the number of polarization channels summed and $\Delta f$ is the bandwidth. A pulse's observed width $W$ may be broadened compared to its intrinsic width $W_i$ by several effects. After dedispersing the raw data and obtaining a dedispersed time series, there is residual dispersive broadening due to the finite width of a frequency channel and the error of the trial dispersion measure (DM) used compared to the actual pulsar DM. Scatter broadening is not correctable and will have a contribution that depends on observing frequency and varies with direction on the sky. In general, 
\be
W \approx \left(W_{\rm i}^2 + \Delta t_{\rm DM,ch}^2 + \Delta t_{\rm DM,err}^2 + \Delta t_{\rm sc}^2\right)^{1/2},
\ee
where $\Delta t_{\rm DM,ch} = 8.3\mu s {\rm DM} \Delta f_{\rm ch}/f_{\rm}^3$ is the dispersive broadening across a frequency channel of width $\Delta f_{\rm ch}$ (MHz) for an observing frequency $f$ (GHz), $\Delta t_{\rm DM,err}$ is the dispersive broadening due to the difference between the trial and actual DM of the source, and $\Delta t_{\rm sc} \propto f^{-4}$ is the scattering broadening. Broadening conserves pulse area, so that the intrinsic and observed peak flux densities are related through $S_{\rm p,i} W_i = S_{\rm p} W$. If $S_{\rm p,min} = m \sigma_{\rm n}$ is the detection threshold, the minimum detectable 
{\em intrinsic} peak flux density is
\be
S_{\rm p,i,min} = \left(\frac{W}{W_i}\right) \frac{m S_{\rm sys}}{\sqrt{N_{\rm pol}~\Delta f~W}}.
\ee
For a one steradian pulsar radio beam, a source of intrinsic peak luminosity $L_{\rm p,i}$ can be detected out to a maximum distance of
\be
D_{\rm max} = \left(\frac{L_{\rm p,i}}{S_{\rm p,i,min}}\right)^{1/2} = L_{\rm p,i}^{1/2} \left(\frac{W_{\rm i}}{W}\right)^{1/2} \frac{\left(N_{\rm pol}~\Delta f~W\right)^{1/4}}{\left(m S_{\rm sys}\right)^{1/2}}.
\ee
For a steadily emitting pulsar with period-averaged luminosity $L$ and duty cycle $f_{\rm dc}$, we have $L_{\rm p} \approx L/f_{\rm dc}$.

\begin{figure}[h!]
\centering
\includegraphics[scale=0.4]{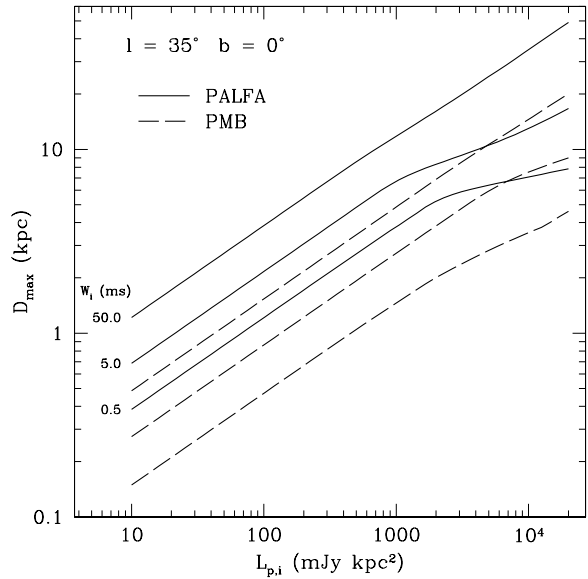}
\caption{Maximum distance at which transients with various peak 
luminosities and intrinsic pulse widths of (top to bottom for each set of curves) 50, 5, and 0.5~ms 
can be detected by the PALFA and Parkes Multibeam surveys. The linear 
portion of the curves corresponds to a luminosity-limited regime. 
In that regime, $D_{\rm max}$ is about twice as large for PALFA than 
for the Parkes Multibeam survey because of the larger sensitivity of 
the Arecibo telescope. Breaks in the curves correspond to a 
transition from luminosity-limited to scattering-limited detection 
for PALFA and to a dispersion-limited regime for Parkes. 
The difference is due to the smaller channel width of PALFA (0.4~MHz) 
compared to Parkes (3~MHz).}
\label{fig:dmax_vs_lp}
\end{figure}

The amount of pulse broadening depends on system parameters as 
well as dispersion and scattering, which vary with direction on 
the sky so that 
$W = W(W_{\rm i} ,l,b,f,\Delta f,N_{\rm ch},S_{\rm sys},\Delta t)$. 
We use the NE2001 model of Galactic ionized electron density 
(\citealt{NE2001}) to calculate representative results for 
$D_{\rm max}$ in the direction $l = 35\degree, b = 0\degree$, 
a region of overlap between PALFA and the Parkes Multibeam survey. 
Fig.~\ref{fig:dmax_vs_lp} shows $D_{\rm max}$ vs. $L_{\rm p,i}$ 
detection curves using a threshold $m=6$ for both surveys. 
For lower luminosities, sources 
are not visible to large enough distances for scattering to affect 
detectability and the inverse square law dominates the detection 
curve so that $D_{\rm max} \propto L_{\rm p,i}^{1/2}$. For larger 
distances and smaller intrinsic pulse widths, scattering and 
(residual) dispersion smearing  
make pulses increasingly harder to detect and 
$D_{\rm max}$ increases more slowly with $L_{\rm p,i}$. 

\section{Single Pulse Search Methods}\label{section_spmethods}

This section describes processing methods employed by the Cornell pulsar search pipeline\footnote{\texttt{http://arecibo.tc.cornell.edu/PALFA/}} at the Cornell Center for Advanced Computing and the Swinburne University of Technology. The PRESTO search code \footnote{\texttt{http://www.cv.nrao.edu/$\sim$sransom/presto}} is run independently at West Virginia University, University of British Columbia, McGill University and the University of Texas at Brownsville. PRESTO uses a similar matched filtering algorithm to the one described below, but a different RFI excision scheme and trial DM list. 

We dedisperse raw data with 1272 trial DMs in the range $0 - 1000$~pc~cm$^{-3}$. In order to find individual pulses from intermittent sources we operate on dedispersed time series with two time domain algorithms: matched filtering, which has been the standard in single pulse searching so far, and a friends-of-friends algorithm. After single pulse candidates have been identified we use a stacking method in the time-frequency plane to verify that the pulses are dispersed and the sweep observed across frequencies follows the dispersion relation, as expected for non-terrestrial sources. Certain types of terrestrial signals, e.g. swept-frequency radars, have pulses whose appearance in the time-frequency plane may approximate dispersion by ionized gas, but on closer inspection such pulses generally deviate significantly from the cold plasma dispersion relation. 

\subsection{Matched Filtering}

Matched filtering detection of broadened pulses relies on the convolution of a pulse template with the dedispersed time series. Ideally, a pulse template would consist of one or several superimposed Gaussian templates, reflecting the diversity of pulsar pulse profiles, some of which exhibit multiple peaks. In addition, multipath propagation due to scattering adds an exponential tail to the pulse profile. We approximate true matched filtering by smoothing the time series by adding up to $2^{\rm n}$ neighboring samples, where $n = 0-7$, and selecting events above a threshold after each smoothing iteration. The smoothing is done in pairs of samples at each $n$-stage, so that the resulting template is a boxcar of length $2^{\rm n}$ samples (\citealt{JimMaura03}).
The sampling time used in PALFA observations is 64~$\mu$s and therefore our matched filtering search is most sensitive to pulse widths of 64~$\mu$s to 8.2~ms. This search strategy is not optimal for single pulses from heavily scattered pulsars because the templates are symmetric, while the scattered pulse shape with an exponential tail is not, and significantly scattered pulses can be wider than $\sim 10$~ms.

\subsection{Finding Time-Domain Clusters}

Two issues make a complement to matched filtering necessary. 
The pulse templates described above have discrete widths of $2^n$ 
samples by algorithm design and there is decreased sensitivity to pulses 
with widths that are significantly different. In addition, 
the matched filter search output may be dominated by bright, wide pulses 
from radio frequency interference (RFI). In that case, a single 
RFI burst is detected as an overwhelming number of individual events instead 
of a single event. The cluster algorithm is similar to the friends-of-friends search algorithm used to find galaxies in optical images
(\citealt{Huchra82}), and complements matched filtering by not restricting 
the width of expected pulse detections. The dedispersed time series 
is processed sequentially. Each event above a threshold which is found 
 is designated as the first of a cluster. A cluster of events 
is augmented by broadening it to include any adjacent samples 
above a threshold. Gaps of $n_{\rm gap}$ samples are allowed within a cluster, and PALFA data is processed with $n_{\rm gap} = 2$. 
The brightest sample of a cluster is recorded as the event 
amplitude, and the total number of samples in the cluster as its width. 
This approach is less sensitive to weak, narrow pulses but results in 
significantly fewer spurious events due to RFI.

A limiting factor for the largest pulse width detectable by both 
the matched filtering and friends-of-friends search algorithms is 
the fact that data are analysed in blocks much shorter than the complete 
time series length. The mean and standard deviation of a block 
are used for thresholding events found within the block. This 
approach minimizes the effect of baseline variations with time scales 
much longer than typical pulsar pulse widths of a few to a few tens of ms. 
The disadvantage is that only pulses with $W \ll T_{\rm block}$ can 
be detected. In processing PALFA data, we use blocks of length 
$T_{\rm block} = 4096 \times 64~\mu$s~$= 0.26$~s. According to the 
NE2001 model of ionized gas in the Galaxy (\citealt{NE2001}), 
a scattering broadening time of that magnitude in the inner-Galaxy part 
of the PALFA survey corresponds to $\rm DM > 1000$~pc~cm$^{-3}$ and 
a maximum search distance well outside of the Milky Way for 
most directions we survey. However, lines of sight intersecting HII 
regions can result in large dispersion measures and scattering 
times and are therefore selected against.  Any intergalactic scattering
that broadens pulses beyond about 0.1~s will also cause events to be
selected against.  

\subsection{Example Results}

\begin{figure}
\centering
\includegraphics[scale=0.4]{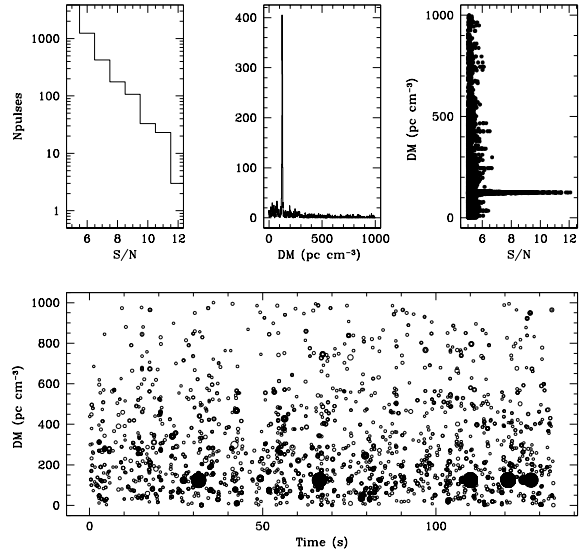}
\caption{Single pulse search output for PSR~J0627+16. The bottom panel shows events with $\rm S/N > 5$ vs. time and DM with larger circles denoting brighter bursts. Panels on top from left to right show histograms of the number of events vs. S/N and DM, and a scatter plot of event DM vs. S/N. A fit to the narrow peak of event DM vs. S/N indicates a pulse width $\sim 2$~ms (\citealt{JimMaura03}).}\label{J0627best}
\end{figure}

Fig.~\ref{J0627best} shows standard single pulse search output for the discovery observation of PSR~J0627+16 (\S~8.1). The main panel shows signal-to-noise ratio (S/N) vs. DM and time for events with $\rm S/N > 5$. The panels on top illustrate event statistics for the observation: from left to right, number of events vs. S/N and DM, and DM vs. S/N. The pulsar detection is manifested as a peak in the histogram of number of events vs. DM. There is a corresponding peak in the DM vs. S/N plot, since the several detected pulses are significantly brighter than background noise in the dedispersed time series. 

\begin{figure}
\centering
\subfigure{\includegraphics[scale=0.4]{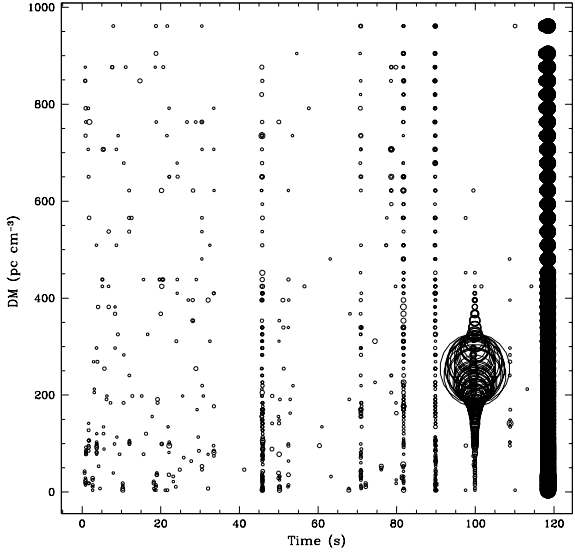}}
\subfigure{\includegraphics[scale=0.4]{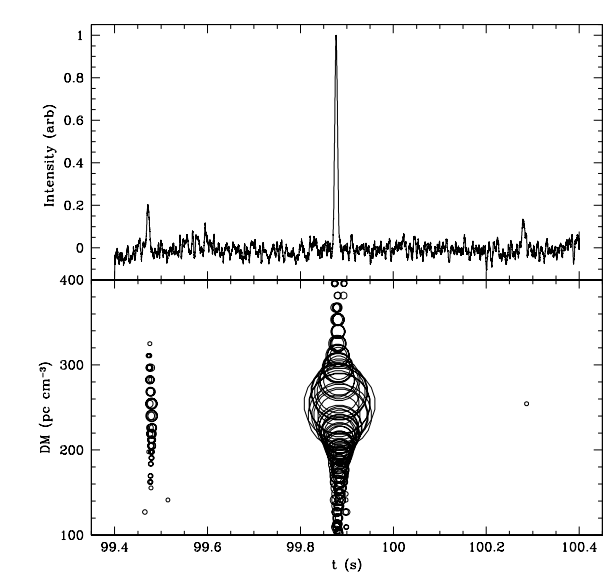}}
\caption{In the upper panel, events with $S/N > 4$ from the discovery PALFA observation of PSR~J1928+15 are displayed in the DM-time plane. Events aligned vertically and spanning all trial DM values at $t \sim 46, 71, 82, 90, 118$~s are due to terrestrial interference. Three closely spaced bursts were found at t $\sim$ 100~s, DM $\sim$ 250~pc cm$^{-3}$. The lower panel shows a magnification of the area around the bursts in the dedispersed time series (top) and the DM-time plane (bottom). The intervals between successive bursts are 0.403~s, establishing the pulsar period. Larger circles denote brighter bursts, and the brightest burst has $S/N \sim 60$ in the dedispersed time series. The scaling of circle size with signal-to-noise ratio is slightly different in the two plots.}\label{rratquick}
\end{figure}

Fig.~\ref{rratquick} shows events in DM-time space for another PALFA single pulse discovery, PSR~J1928+15 (\S~8.5). In this case, three closely spaced bursts were found at $t \sim 100$~s, $\rm DM \sim 250$~pc~cm$^{-3}$. The pulsar is detected at a range of DMs, with the signal to noise ratio increasing as trial DM values approach the actual pulsar DM and decreasing as trial DMs further on recede from the pulsar DM. In contrast, events due to interference from terrestrial signals at $t \sim 46, 71, 82, 90, 118$~s span the entire range of trial DMs and their signal to noise ratios do not show a significant variation with DM. 

\begin{figure}
\centering
\includegraphics[scale=0.4]{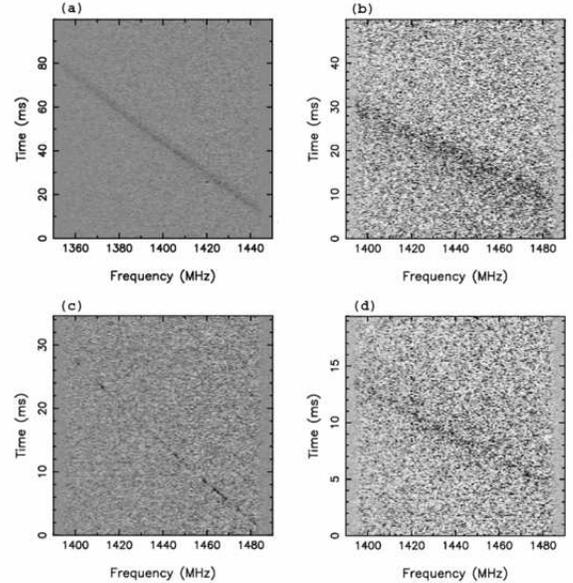}
\caption{Dispersion sweep in the time-frequency plane of the 
brightest single pulse detected from (a) PSR~J1928+15 and (b) PSR~J1946+24; 
(c) Stacked dynamic spectrum of the five brightest pulses 
detected from PSR~J0627+16; 
(d) Stacked dynamic spectrum of the two brightest pulses 
detected from PSR~J1909+06. 
The pulses are dispersed, such that the higher 
frequencies arrive earlier.}
\label{4dynspecs}
\end{figure}

If an excess of candidate pulses is identified in the dedispersed time series for a particular trial DM, we use the expected dispersion sweep across the frequency band in order to test if the pulses are from non-terrestrial origin. For a pulse with $\rm S/N \gtrsim 10$ in the time series, we extract a chunk of raw data centered on its time of arrival and look for a sweep across frequency that follows the dispersion relation (Fig.~\ref{4dynspecs} a, b). 

\subsection{Time-Frequency Plane Stacking}

In the case of multiple weaker pulses detected in the same beam, we extract a raw data chunk centered on the arrival time of each pulse, stack the chunks and look for a dispersion sweep in the resulting cumulative dynamic spectrum (Fig.~\ref{4dynspecs} c, d). 

Stacking can induce spurious dispersion sweeps in the time-frequency plane
even when the data consist of only Gaussian-distributed noise. 
Single pulse events identified in dedispersed time series by definition
correspond to sums over a given dispersion path that are above average.  When these 
are used to select chunks for summing, the statistical
fluctuations will build up to show a dispersed pulse in the time-frequency
plane  that follows the
cold-plasma dispersion law perfectly. 
We simulated 
this effect by generating a time-frequency plane of Gaussian noise, 
dedispersing it, and selecting events above threshold from 
the resulting time series. We extracted chunks from the fake 
time-frequency plane centered on each event and stacked them, 
thus reproducing the procedure used on real data. 

Selecting events with signal-to-noise above a threshold $m_d$ in the dedispersed time series yields a biased set of noise-only samples in the time-frequency plane. On the other hand, when stacking dynamic spectra, points in the stacked dynamic spectrum must have a minimum signal-to-noise $m_c \sim 2$ for a dispersion sweep to be discernible by eye. The rms noise in the dedispersed time series after summing $N_{\rm ch}$ channels is a factor of $\sqrt{N_{\rm ch}}$ larger than the rms noise in the time-frequency plane. The rms noise in the stacked dynamic spectrum after stacking $N_{\rm c}$ chunks is a factor of $\sqrt{N_{\rm c}}$ larger than the rms noise in the time-frequency plane. We calculate the average deviation from the mean in the time-frequency plane implied by the two thresholds $m_d$ and $m_c$, equate them, and obtain
\be
N_{\rm c} \gtrsim \left(m_{\rm c} / m_{\rm d}\right)^{2}N_{\rm ch}.
\ee

For the PALFA survey and our processing pipeline, 
$N_{\rm ch} = 256$ and $m_d = 5$. Therefore 
if $N_{\rm c} \gtrsim 40$ data chunks are added
there  will be a spurious, dispersed  pulse in the resulting stacked dynamic spectrum even if the chunks contain only noise.
Since the noise decorrelates over $1-2$ samples, the pulse will have a 
width on the order of $64-128~\mu$s. 
To avoid the induced spurious-event
effect, the width and S/N of stacked dynamic spectra should be compared
with those that would result from noise only. All of our detections that use stacking satisfy the criteria $W >> 64~\mu$s and $N_{\rm c} << 40$. Typically we sum chunks centered on the brightest $2-5$ pulses for a single pulse candidate, which is safely below the $N_c$ limit. In the case of PSR~J1909+06 
(Fig.~\ref{4dynspecs}d) only 2 chunks were added and the dispersed 
swath in the time-frequency plane is 1~ms wide. 

\subsection{RFI Excision}

The RFI environment at the Arecibo telescope and the 7-beam configuration of the ALFA receiver present both challenges and opportunities for RFI mitigation to facilitate searching for single pulses. PALFA survey data are dedispersed with trial dispersion measures of $0 - 1000$~pc~cm$^{-3}$. RFI pulse intensity typically peaks at $\rm DM = 0$~pc~cm$^{-3}$, and incidental low-intensity pulses whose S/N peaks at DM = 0~pc~cm$^{-3}$ tend to be smeared below the detection threshold for dedispersed time series with trial $\rm DM > 50-100$~pc~cm$^{-3}$. A more complex signature is observed for Federal Aviation Administration (San Juan airport) radar pulses, which are unfortunately common in Arecibo observations. The radar rotation period is $P_r = 12$~s, and each pulse has an envelope that is $\sim 1$~s wide and consists of sub-pulses with variable period on the order of $2-3$~ms. Depending on telescope orientation, radar pulses may be detected in all, some, or none of the ALFA beams due to reflections off the telescope structure. Radar pulses are up to two orders of magnitude brighter than pulsar pulses and, without mitigation, can completely dominate single pulse search results. In addition, the modulation of the radar signal is manifested as detections with $\rm DM \neq 0$~pc~cm$^{-3}$ so that unlike other non-radar RFI, their S/N vs. DM signature cannot be used for excision.

We exploit the known radar characteristics as well as the pattern of pulse detection in the 7 ALFA beams to excise both radar and non-radar RFI. The first part of our excision algorithm targets radar pulses. After a list of single pulse events is generated for all trial DMs, we bin the events for trial $\rm DM = 0-3$~pc~cm$^{-3}$ in time (by 0.1~s) and record the number of events in each time bin. Then we treat the histogram as a time series and perform a discrete Fourier transform. A peak near the radar rotation frequency indicates a significant number of radar pulses in the data. From the Fourier components we extract the phase and find the arrival time of the earliest radar pulse, $t_0$. Since the envelope width is $\sim 1$~s, events within 0.5~s of that location are excised, and the procedure is repeated for events near $t = t_0 + N P_r$, where $N$ is an integer. The more radar pulses are present, the better the performance of this technique because the radar peak in the DFT is more prominent and the pulses' arrival times are determined more precisely. However, if only a couple of strong radar pulses are present within the typical 268~s PALFA integration time, they may be bright enough to dominate event statistics, yet the DFT method does not excise non-periodic incidental RFI. Consequently, after applying the DFT-based method we use an additional RFI filter that handles aperiodic cases. 

\begin{figure}
\begin{minipage}[b]{1\linewidth}
\centering
\includegraphics[scale=0.4]{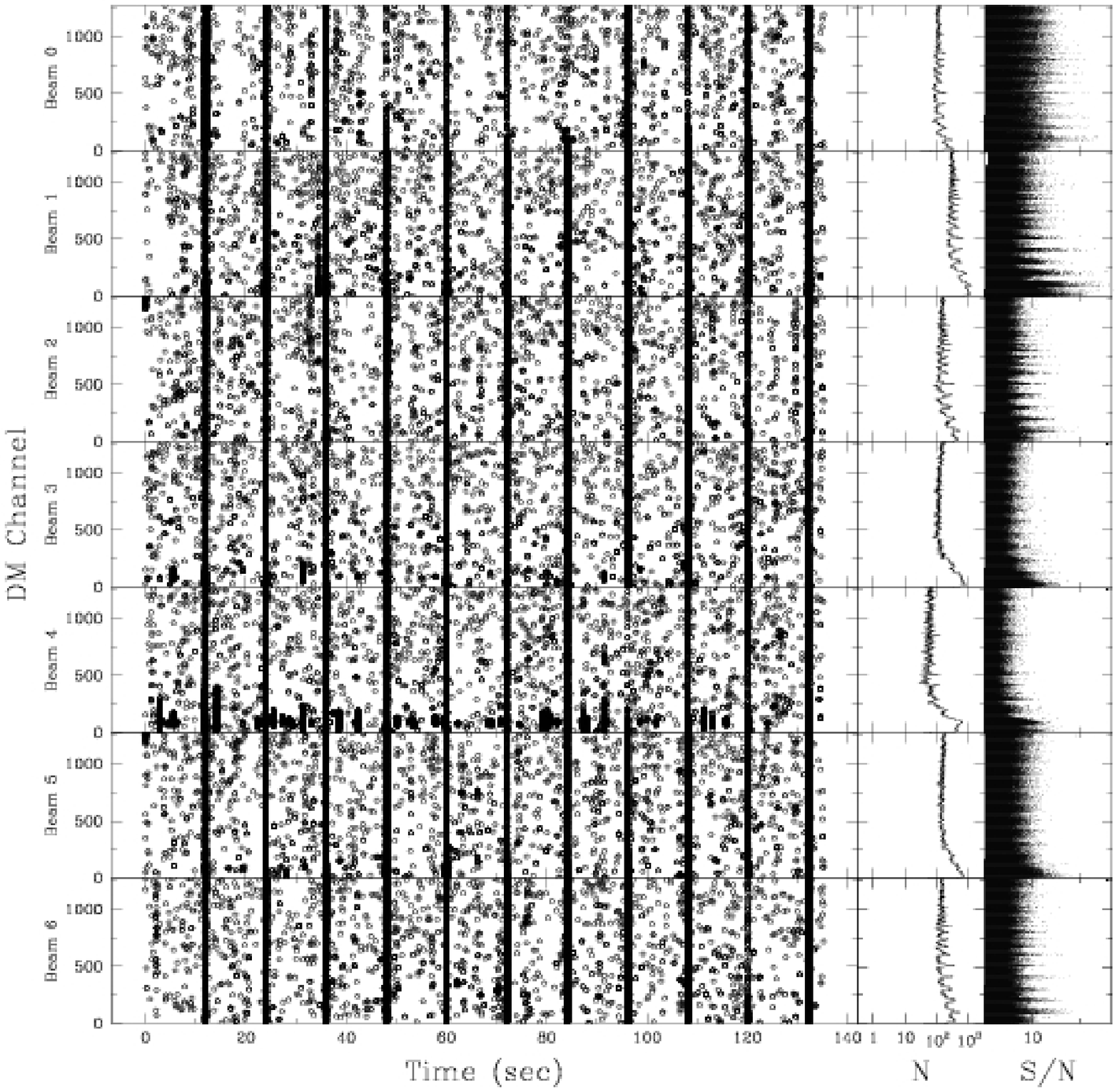}
\end{minipage}
\begin{minipage}[b]{1\linewidth}
\centering
\includegraphics[scale=0.4]{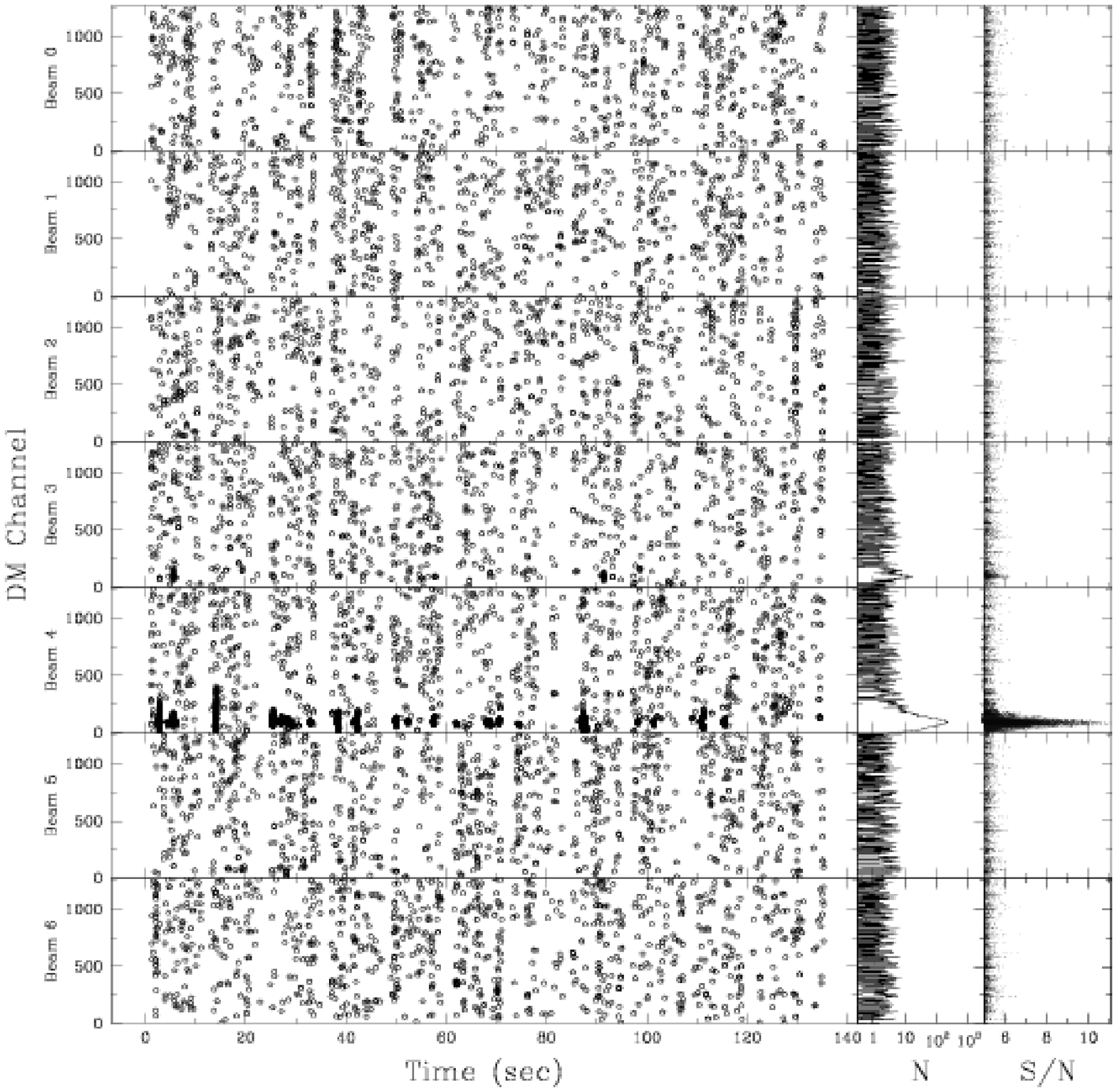}
\end{minipage}
\caption{Single pulse search output for a blind detection of pulsar B2020+28 before (top) and after (bottom) excising radar and incidental RFI. Each row shows results from one ALFA beam. From left to right, panels show: events with $S/N > 5$ vs. DM channel and time, number of events vs. DM channel, and event S/N vs DM channel. The pulsar detection in beams 3 and 4 is evident after RFI events are excised.}\label{SPcleaned}
\end{figure}

The second filter uses the number and proximity of beams in which an event is detected in order to determine if it is due to RFI. Again events for trial $\rm DM = 0-3$~pc~cm$^{-3}$ from each beam are binned in time. After detecting peaks indicating an excess of events for the respective time bins, the algorithm cross-checks between results for all seven beams, and each event falling within a histogram peak receives a penalty grade based on how many beams' histograms exhibit a peak and how close to each other they are on the sky. Most pulsars detected blindly via a single pulse search appear in one beam or two adjacent beams, and very bright pulsars may be detected in three or four adjacent beams. We set the excision penalty threshold just below the value corresponding to the latter configurations and excise events accordingly. This method complements the DFT cleaning scheme and effectively excises sparse radar blasts as well as non-periodic RFI detected in multiple beams. The application of the two excision algorithms makes a marked difference in the final single pulse search output for pointings contaminated with RFI (Fig.~\ref{SPcleaned}). The figure shows some false positives, for example a clump of pulsar pulses in beam 4 around $t = 80$~s are excised due to low-level RFI at low DMs occurring in non-adjacent beams in the same 0.1~s time bins as the pulses. Lowering the threshold according to which an excess of events is defined and time bins are marked for excision reduces false positives but also diminishes the effectiveness of RFI excision. While tens of pulses are detected within 134~s from the bright pulsar B2020+28 (Fig.~\ref{SPcleaned}), for sources discovered via a single pulse search the number is an order of magnitude smaller (see \S~7). Therefore the chance of RFI occurring simultaneously with pulses of intermittent sources and causing them to be excised is much lower. 

\section{PALFA Pulsar Detection Statistics}\label{section_palfastats}

The PALFA survey has made a total of 354 blind detections of 
172 pulsars up to late 2008. Following quick-look  
processing of the data 
with degraded time and frequency resolutions 
at the time that data are acquired (\citealt{palfa1}),
data are shipped to external sites for processing through the 
Cornell pulsar search code and PRESTO. 
Table~\ref{detection_stats} shows a breakdown of Cornell detections, including pulsars originally discovered by the quick-look pipeline, by 
the periodicity (FFT) and single pulse (SP) search algorithms for 
known pulsars and new discoveries, and 
Table~\ref{psr_params} lists parameters for PALFA single pulse 
discoveries that are discussed in more detail in \S~\ref{section_objects}.
Columns in Table~\ref{psr_params} are
(1) Pulsar name;
(2)-(3): equatorial coordinates; 
(4) period; 
(5) pulse width (FWHM);
(6) dispersion measure; 
(7) distance; 
(8) peak flux density of the brightest detected pulse; 
(9) total observation time;
(10) total number of pulses detected;
and
(11) implied average pulse rate.
The last column (12) points to the paper section that discusses the object. 
 


\begin{deluxetable}{ccccc}
\tablecolumns{5}
\tablewidth{0pc}
\tablecaption{Pulsar detection statistics by algorithm.\label{detection_stats}}
\tablehead{\colhead{Pulsars} &\colhead{ FFT only} & \colhead{SP only} & \colhead{FFT and SP} & \colhead{Total}}
\startdata
Known & 48 (38\%) & 5 (4\%) & 73 (58\%) & 126 \\
New & 28 (61\%) & 6 (13\%) & 12 (26\%) & 46
\enddata
\end{deluxetable}


\begin{deluxetable*}{lllcccccrrrc}
\tablecolumns{12}
\tablewidth{0pc}
\tablecaption{Parameters of PALFA single pulse discoveries
\label{psr_params}}
\tablehead{\colhead{Pulsar} & \colhead{$RA$~\tablenotemark{a}} & \colhead{$DEC$~\tablenotemark{a}} & \colhead{$P$} & \colhead{$W$} & \colhead{DM} & \colhead{$D$~\tablenotemark{b}} & \colhead{$S_{\rm p}$~\tablenotemark{c}} & \colhead{$T_{\rm tot}$} & \colhead{$N_{\rm tot}$} & \colhead{Rate} & \colhead{Comment}\\ 
 & (hh:mm:ss) & (dd:mm) & (s) & (ms) & (pc~cm$^{-3}$) & (kpc) & (mJy) & (s) &  & (h$^{-1}$)}
\startdata
J0627+16 & 06:27:13(7) &  16:12(2) & 2.180 &	0.3 & 113 &	3.2 &	150 & 7454 & 48 & 23 & \S~\ref{0627sec}\\
J0628+09 & 06:28:33(7) & 09:09(2) & 1.241 &	\phn 10 & \phn 88 &	2.5 & \phn 85  & 1072 & 42 & 141 & \S~\ref{0628sec}\\
J1854+03 &  18:54:09(7) & 03:04(2) &	4.559 &	\phn 50 & 216 &	5.5 & \phn 14 & 388 & 9 &  84 & \S~\ref{1854sec}\\
J1909+06 & 19:09:24(7) & 06:40(2) & 0.741 &	1.5 & \phn 35 &	2.2 & \phn 82  & 536 & 10 & 67 & \S~\ref{1909sec}\\
J1919+17 & 19:19:47(7) & 17:44(2) & 2.081 &	100 & 148 &	5.3 & \phn 12   & 393  & 35 & 320 &  \S~\ref{1919sec}\\
J1928+15 & 19:28:20(7) & 15:13(2) & 0.403 &	\phn \phn 5 & 242 &	7.4 &	180  & 2880 & 3 & 4 & \S~\ref{1928sec}\\
J1946+24 & 19:46:00(7) & 23:58(2) & 4.729 &	\phn \phn 4 & \phn 96 &	4.3 &	101  & 268 & 4 & 54 & \S~\ref{1946sec}\\
\enddata
\tablenotetext{a}{ Position uncertainties correspond to the angular radius (out to 50\% of boresight power) of an individual ALFA beam in the discovery observation.}
\tablenotetext{b}{ Estimate from the NE2001 model of Galactic eletron density (\citealt{NE2001}).}
\tablenotetext{c}{ Peak flux density at 1.4~GHz defined as $S = (S/N)S_{\rm sys}/\sqrt{N_{\rm pol}\Delta f W}$, where $S/N$ is the signal to noise ratio of the brightest detected pulse.}
\end{deluxetable*}

More than half of the detected known pulsars and only a quarter of the newly discovered objects were seen by both the periodicity and single pulse search algorithms. There is also a significant difference in the fraction of pulsars detected only via a single pulse search. In total, 74\% of the new pulsars as opposed to 42\% of the known pulsars were detected either only by FFT search or only by single pulse search. The much higher percentage of new pulsars to be detected only by one algorithm means that most of the PALFA discoveries are either periodic emitters too weak to show up in single pulse searches or intermittent objects whose emission is too heavily modulated be detected by a periodicity search. Considering that bright pulsars are more likely to be seen by both the FFT and SP search, and they are also more likely to have already been found by previous, less sensitive surveys, this indicates that PALFA is probing deeper and finding pulsars that are farther away or less bright than the known objects in the same region of the Galactic plane. 

Single pulse searches have not been routinely carried out on survey data in the past, with the exception of \cite{Phinney79}, \cite{Nice99}, who discovered PSR~J1918+08 via a single pulse search, and \cite{McLaughlin06}, who found 11 RRATs. Therefore, the proportion of single-pulse-only detections is higher for the PALFA discoveries than the known pulsars. The five known pulsars detected only by single pulse search are known steady emitters which were seen away from the beam center and therefore with reduced sensitivity. 


\section{Selection Effects and Intermittency}\label{section_biases}

The sample of PALFA single pulse discoveries (\S~\ref{section_objects}) includes one object which has not been detected again despite reobservations (J1928+15), two sources detectable in re-observations through their time-averaged emission (PSR~J0628+09 and PSR~J1909+06), and four long-period objects, one of which has the smallest known duty cycle for any radio pulsar (PSR~J0627+16). 

Among the 11 Parkes RRATs, 10 were successfully redetected in subsequent Parkes observations, and one (J1839$-$01) has not been redetected despite multiple attempts (\citealt{McLaughlin06}). PSR~J0848$-$43 and PSR~J1754$-$30 can sometimes be detected through their time-averaged emission with sensitive, low-frequency observations (\citealt{McLaughlinRev07}). Six of the sources have periods greater than 2~s. 

In this section we discuss selection effects which may account for some of the observational characteristics of intermittent sources. Characteristics of this population which remain unexplained by observational biases may indicate underlying intrinsic differences between these objects and conventional radio pulsars. They may belong to a genuinely intermittent and physically different class of neutron stars.

\subsection{Multiplicative Effects}

We can define a general model for observed signal intensity in terms of the time and frequency-dependent flux density $S_{\rm i}(t,\nu)$ as
\be
I\left(t,\nu\right) = G_{\rm t}\left(t,\nu\right) G_{\rm ISS}\left(t,\nu\right) S_{\rm i}\left(t,\nu\right) + n\left(t,\nu\right) + {\rm RFI}\left(t,\nu\right),
\ee
where $G_t$ is the telescope gain, $G_{\rm ISS}$ is the variation factor due to interstellar scintillation, $n(t,\nu)$ is radiometer noise, and $RFI(t,\nu)$ is the contribution from terrestrial radio frequency interference. 

\begin{figure}
\centering
\includegraphics[scale=0.4]{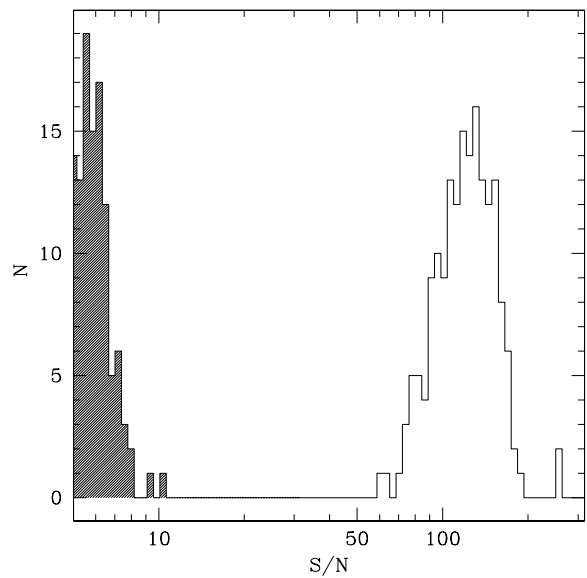}
\caption{Observed single pulse PDFs for PSR~B1933+16 in two beams of a PALFA pointing: on source (right) and 6 arc minutes away (left, shaded). All pulses are detected individually in the on-source beam, while only the high intensity tail of the distribution is detected via a single pulse search in the off-source beam. A sidelobe detection of a canonical pulsar will pick out the brightest pulses and may misrepresent the pulsar as an intermittent source.}\label{pulsepdf}
\end{figure}

Variations in telescope gain across the beams of the multi-beam receiver or within the power pattern of a single beam impact the detectability of a pulsar as either a single pulse or periodic source. As shown by the reobservations of PSR~J0627+16, PSR~J1909+06, PSR~J0848$-$43 and PSR~J1754$-$30, classifying a source as intermittent may be due to insufficient sensitivity to detect its periodic emission during the discovery observation. The effectiveness of single pulse vs. FFT-based periodic search depends on the pulse amplitude distribution, and the observed pulse amplitude distribution in turn depends on the pulsar location with respect to the telescope beam power pattern. Fig.~\ref{pulsepdf} shows the probability density function (PDF) of event amplitude for PSR~B1933+16, a strong pulsar detected in two beams of the same PALFA pointing. One beam of the ALFA receiver is pointing directly at the pulsar and the low-intensity tail of the distribution is clearly visible since all pulses are above the detection threshold. The pulsar is also detected in the first sidelobe of an adjacent beam, in which case only the brightest pulses are above the detection threshold and only the high-intensity tail of the distribution is visible. Thus an off-axis detection of a steadily emitting pulsar may misrepresent it as an intermittent source. 

A different set of effects which affect detectability in single pulse vs. periodic search is due to intervening ionized gas. Turbulence in the ionized interstellar gas and the motion of the pulsar with respect to the gas introduce phase modulations in the pulsar emission which cause the observed intensity to change over time and frequency. In general, the scintillation gain factor has contributions from both diffractive (DISS) and refractive (RISS) interstellar scintillation, such that $G_{\rm ISS}\left(t,\nu\right) = G_{\rm DISS}\left(t,\nu\right) G_{\rm RISS}\left(t,\nu\right)$ \citep{Rickett90}. The detectability of nearby pulsars with low DMs can be significantly affected by DISS, which can have time scales comparable to typical survey pointing times (minutes) and frequency scales $\Delta f_{\rm DISS} << f$. In terms of the observing frequency $f$ and the pulsar distance $D$, the DISS time scale $\Delta t_{\rm DISS} \propto f^{1.2} D^{-0.6}$. The scintillation bandwidth $\Delta f_{\rm DISS}$ and scattering timescale $\tau_{\rm s}$ are related through $2\pi\Delta f_{\rm DISS}\tau_{\rm s} \approx 1$. An empirical fit from \cite{Bhat04} for $f$ in GHz and $\tau_{\rm s}$ in ms gives an estimate for $\tau_{\rm s}$ based on DM:
\be
\log \tau_{\rm s} = -6.46 + 0.154~\log~\rm DM + 1.07\left(\log~\rm DM\right)^2 - 3.86~\log \emph{f} \label{eqn_taus},
\ee
but there is significant scatter about this relationship. When the scintillation bandwidth and time scale are small, averaging over the observation time and bandwidth can quench DISS. 

For refractive scintillation, $\Delta f_{\rm RISS} \propto f$ and $\Delta t_{\rm DISS} \propto f^{0.57} D^{1.6}$. Unlike for DISS, the time scale of RISS increases with distance and therefore it can become very long for distant pulsars with relatively high DMs. Both RISS and DISS can either help or hinder pulsar detection due to the time-variable constructive or destructive interference of wave fronts they cause. For a single-pass survey, this means that some weak pulsars may be detected because they are modulated above the detection threshold, and some pulsars that are otherwise bright enough may be missed because of scintillation modulation. \cite{cl91} discuss in detail the detection probability for scintillating pulsars by single-pass and multi-pass surveys. Since PALFA is a single-pass survey, it may have missed pulsars that scintillate on time scales comparable to or longer than the observation time. None of our discoveries in Table~\ref{psr_params} appear to be affected by DISS, though RISS over long time scales may affect the objects with higher DMs.

\subsection{Observation Time vs Pulsar Period}

The detectability of a pulsar as a periodic source depends on the integrated pulse flux within the observation, which is determined by the number of periods within the observation time, $N_p = T_{\rm obs}/P$, along with the pulse amplitude distribution. For the same integrated flux per pulse, fewer periods within a fixed $T_{\rm obs}$ mean that long-period pulsars are less likely to be detected than short-period pulsars by any method relying on detecting integrated flux such as an FFT-based search or a continuum imaging survey. For FFT searches in particular, this effect is compounded by the fact that when $N_{\rm p}$ is small, the harmonics of interest are in the low-frequency part of the power spectrum, which is often dominated by non-Gaussian red noise and RFI. Hereafter we use the term ``small-$N_{\rm p}$ bias'' to refer to the combined influence of these effects on pulsar detectability.

\section{Intermittency Measure}\label{section_intermittency} 

Both the periodicity and single pulse searches perform with varying efficiency depending on an object's degree of intermittency. In this section we present an intermittency measure method to quantify the relative performance of the two search algorithms, apply it to results from the Parkes Multibeam and PALFA surveys, and discuss general implications for surveys. 

\subsection{Definition}

We can compare the detectability of objects by periodicity and single pulse search and attempt to narrow down different classes of intermittent emitters by calculating the intermittency ratio
\be
r = \left(S/N\right)_{\rm SP}/\left(S/N\right)_{\rm FFT}
\ee
for each object in our sample of PALFA and Parkes detections. \cite{McLaughlin03} derive
\be
r = \left(\frac{2\eta}{\zeta N_{\rm p}^{1/2}}\right)\frac{S_{\rm max}}{S'_{\rm av}}\label{eqn_r},
\ee
where $\zeta \approx 1.06$ and $\eta\sim 1$ for a Gaussian pulse shape, $S_{\rm max}$ is the maximum expected pulse intensity within $N_{\rm p} = T_{\rm obs}/P$ periods and $S'_{\rm av}$ is a modified average pulse peak intensity. 

\begin{figure}[htbp]
\centering
\includegraphics[scale=0.4]{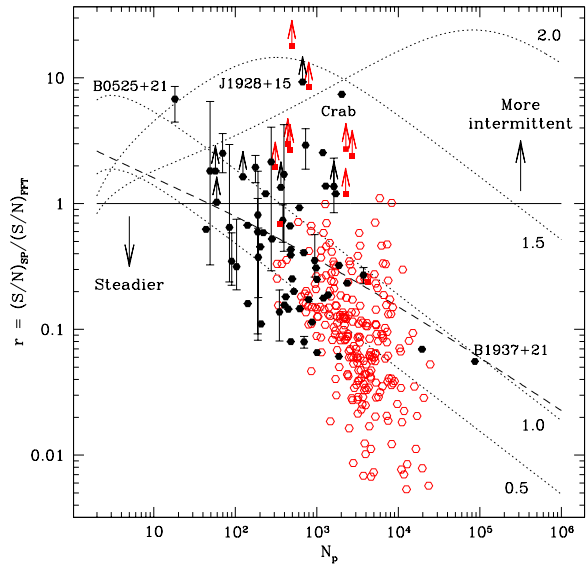}
\caption{The intermittency ratio $r$ vs $N_p = T_{\rm obs}/P$ for 283 PALFA (filled hexagons) and 255 Parkes Multibeam (squares, open hexagons) pulsar detections. Squares denote the 10 Parkes RRATs with period measurements, including two which can sometimes be detected through their time-averaged emission. The criterion for inclusion of canonical pulsars was a simultaneous periodic and single pulse detection. PALFA points include both blind survey detections and targeted test observations of known pulsars. Arrows denote lower limits for PALFA and Parkes intermittent sources. Points with bars show the average $r$ for a set of observations of the same object, with the bars denoting the maximum and minimum $r$ values. Millisecond pulsar detections appear at lower right. Dotted lines show $r$ for power-law pulse amplitude PDFs with indices from 0.5 to 2.0. A ratio of $10^5$ was assumed for the cutoff intensities $S_1$ and $S_2$ (\citealt{McLaughlin03}). The dashed line shows $r$ for an exponential PDF. A single pulse search performs better than an FFT periodicity search for $r > 1$.}

\label{r_vs_np}
\end{figure}

The intermittency ratio is a system-independent measure of the efficiency of the two types of search for objects with different degrees of intermittency. In Fig.~\ref{r_vs_np}, expected values of $r$ vs. $N_p$ are shown for an exponential pulse amplitude PDF and power-law distributions with various indices. Lower limits on $r$ are given for intermittent sources which were not detected via periodicity search. In addition, we show $r$ values for pulsars that were detected using both the periodicity and single-pulse search algorithms in the PALFA and Parkes Multibeam surveys. A total of 283 PALFA pulsar detections and 255 Parkes Multibeam detections are shown in the plot, including multiple detections of some objects. 

\subsection{Implications for Surveys}

In Fig.~\ref{r_vs_np}, long-period pulsars are predominantly found in the upper left (small $N_p$, large $r$) and millisecond pulsars in the lower right (large $N_p$, small $r$).PSR~B0525+21, with $r \sim 7$, has $P = 3.7$~s and $\rm DM = 51$~pc~cm$^{-3}$, which make it susceptible to both the small-$N_p$ bias and diffractive scintillation and therefore it is detected more readily as a single pulse source. At the other extreme is PSR~B1937+21, a millisecond pulsar which emits giant pulses but is nevertheless detected with a much higher S/N in the periodicity search than in the single pulse search ($r < 0.07$) because its period is small compared to the PALFA observation time and the normal pulse amplitude is exceptionally steady (\citealt{Jenet04}). This indicates that the effect of $N_p$ on detectability can overshadow individual properties, and in survey mode millisecond pulsars will show up overwhelmingly as periodic sources. In that case, processing the data with a single pulse search algorithm can reveal the presence of giant pulses in an otherwise steady emitter. 

In-between PSR~B0525+21 and PSR~B1937+21, from upper left to lower right is a relatively smooth distribution of pulsars whose intermittency ratios are predominantly determined by the influence of $N_p$ on the detectability of all types of periodic emitters. We look for unusual objects among the outliers from that trend. There are a number of pulsars detected both in periodicity and single pulse search in the region where $r = 1 - 5$, including PSR~B0656+14, which occasionally emits bursts much brighter than the average pulse and would arguably be classified as a RRAT if located farther away (\citealt{Weltevrede06}). Interspersed among them in the region are most of the PALFA and Parkes intermittent sources. PSR~J1909+06, which was not detected in a periodicity search but is visible as a periodic source when folded with its known period, has $r = 1.3$, and the long-period PALFA intermittent sources PSR~J1854+03 and PSR~J1946+24 have $r = 1.0$ and 1.8, respectively. Of the Parkes intermittent sources, seven have $r < 3$. PSR~J1754$-$30 ($r = 0.2$) and PSR~J0848$-$43 ($r = 0.7$) were detected through their time-averaged emission in low frequency follow-up observations. In the $r > 5$ region we find only five objects: PSR~B0525+21, the Crab pulsar, Parkes intermittent sources PSR~J1317$-$5759 and PSR~J1819$-$1458, and PALFA intermittent source PSR~J1928+15. The Crab pulsar has a high intermittency ratio due to its giant pulses but is otherwise a relatively steady emitter as periodic emission is consistently detected even if the giant pulses are ignored. Of the remaining three intermittent sources none are young pulsars like the Crab. The pulse amplitude distributions of PSR~J1317$-$5759 and PSR~J1819$-$1458 are described by power laws with indices $\sim 1$ (\citealt{McLaughlin06}), smaller than measured giant pulse indices of $\sim 2 - 3$ (e.g. \citealt{kt00}, \citealt{c+04}).

We conclude from Fig.~\ref{r_vs_np} is that periodicity and single pulse searches should be used in combination by pulsar surveys since there are types of sources which are best detected by either method. As more intermittent objects are found, placing them in $r - N_{\rm p}$ space may help us identify physically distinct populations. In addition, repeated sky coverage can probe intermittency at time scales of days to months.

\section{Single Pulse Discoveries}\label{section_objects}

Table~\ref{psr_params} lists parameters of the seven pulsars discovered by PALFA via a single pulse search. In this section we discuss each object's properties in detail and describe the steps taken in addition to the PALFA processing pipeline in order to verify the signals' celestial origin and calculate the pulsar period from times of arrival of single pulses. 

\subsection{PSR~J0627+16}\label{0627sec}

PSR J0627+16 was discovered when nine single pulses were detected at a trial $\rm DM = 125$~pc~cm$^{-3}$ (Fig. \ref{J0627best}). The period estimated from the pulse arrival times in the discovery observation is $P = 2.180$~s. What is unusual about this object is its very narrow peak in S/N vs. DM space, indicating a narrow pulse (\citealt{JimMaura03}). 
Fig.~\ref{4dynspecs}c shows the cumulative dynamic spectrum after stacking raw data chunks aligned around the five brightest pulses. A fit to the dispersion sweep in the time-frequency plane gives a best DM of 113~pc~cm$^{-3}$ and when the raw data chunk around the brightest pulse is dedispersed with this DM, the FWHM pulse width is only 0.3~ms. 

Two follow-up observations at 0.33 and 1.4~GHz yielded 17 pulses in 50 minutes and 22 pulses in 72 minutes, respectively. After removing bright single pulses from the dedispersed time series and replacing the respective time samples with a running average, periodic emission with $P = 2.180$~s was detected throughout the 0.33~GHz observation, confirming the period estimate from the discovery observation at 1.4~GHz and the presence of underlying normal emission. 



Many of the RRATs detected by \cite{McLaughlin06} have short duty cycles, calculated as the ratio between the average width of single pulses and the period. PSR~J0627+16 has the smallest known duty cycle, $f_{\rm dc} = 0.01\%$ by this definition. However, for steady emitters the duty cycle is typically defined as the ratio of the folded pulse profile FWHM and the period. Individual pulses may occur in a phase window wider than any individual pulse and corresponding to the folded profile width. At 1.4~GHz, the individual 0.3~ms pulses of J0627+16 occur in a 8~ms window, while at 0.33~GHz individual pulses have widths of $0.3 - 2$~ms and occur in a 15~ms window. The folded pulse profile at 0.33~GHz has FWHM of 60~ms, suggesting that bright pulses may be confined to a narrower window than pulses comprising the underlying normal emission. 

The detection of bright single pulses and a weak periodic signal from PSR~J0627+16 when data are folded with an accurate period estimate, along with the absence of a periodic detection in search mode weighs in favor of the proposition of \cite{Weltevrede06} that some RRATs may have periodic emission with highly variable pulse amplitudes. 



\subsection{PSR~J0628+09}\label{0628sec}

PSR J0628+09 was discovered by detecting three pulses at $\rm DM = 88$~pc~cm$^{-3}$. The period estimated from the pulse times of arrival was 2.48~s. In follow-up observations the pulsar was also detected in periodicity searches as it emits on average several bursts per minute. The periodicity detections and the larger number of single pulses in subsequent observations allowed the actual pulsar period of 1.241~s to be determined (\citealt{palfa1}).


\subsection{PSR~J1854+03}\label{1854sec}

PSR J1854+03 was discovered in 2008 via a single pulse search performed on full-resolution survey data. Four pulses were detected at $\rm DM = 216$~pc~cm$^{-3}$ during the 268~s observation. The period $P = 4.559$~s was estimated by taking the smallest difference between times of arrival (TOAs) of two consecutive pulses and verifying that intervals between all four TOAs are integer multiples of it. A confirmation observation with the more sensitive central ALFA beam yielded within 120~s five pulses whose arrival times match the estimated period. 

\subsection{PSR~J1909+06}\label{1909sec}

PSR J1909+06 was observed in 2006 but not identified as a candidate until 2007 when the full-resolution data were searched for single pulses. Two pulses with a width of $\sim 1$~ms and signal to noise ratio of 6 and 9 were detected at $\rm DM = 35$~pc~cm$^{-3}$. 
The stacked dynamic spectrum clearly shows a dispersion sweep (Fig.~\ref{4dynspecs}d). PSR~J1909+06 was discovered in data from the off-axis beam 2 of the multi-beam ALFA receiver. Since the on-axis gain of the center beam is 10.4~K/Jy compared to 8.2~K/Jy for the other six beams (\citealt{palfa1}), we aimed the center beam at the discovery coordinates for a confirmation observation. For the same integration time of 268~s, 8 pulses with $S/N > 5$ were detected in the confirmation observation. We used the pulse arrival times to determine a period and arrived at a best estimate of $P = 0.741$~s. 

\subsection{PSR~J1919+17}\label{1919sec}

During the 2007 discovery observation of PSR~J1919+17, multiple pulses at $\rm DM = 148$~pc~cm$^{-3}$ were detected in beam 4 of the ALFA receiver, with no corresponding periodic detection. A Fast Folding Analysis (\citealt{Staelin68}, \citealt{Kondratiev09}) was used to narrow down the period to $P = 2.081$~s, which was confirmed in 2008, when the pulsar was detected as a normal periodic emitter with the more sensitive central ALFA beam. The pulsar has a double-peak profile with the two peaks $\sim 100$~ms apart, which most likely made it difficult to find its period from pulse arrival times alone. 


\subsection{PSR~J1928+15}\label{1928sec}

PSR J1928+15 was discovered in 2005 by detection of what looked like a single bright pulse at $\rm DM = 245$~pc~cm$^{-3}$ in a 120~s observation (Fig. \ref{rratquick}, top). More detailed analysis revealed that the event was in fact composed of 3 separate pulses occurring at intervals of 0.403~s, with the middle pulse being brighter by an order of magnitude than the other two (Fig. \ref{rratquick}, bottom). In Fig. \ref{4dynspecs}a the dispersion of the brightest pulse by ionized interstellar gas is shown in the time-frequency plane, evidence of the non-terrestrial origin of the pulses. A fit to the pulse signal in the time-frequency plane resulted in a refined estimate of $\rm DM = 242$~pc~cm$^{-3}$. Despite multiple follow-up observations, the source has not been detected again. 

Given the DM of this source, 242~pc~cm$^{-3}$, it is unlikely that the non-detection can be attributed to diffractive scintillation. Since the three pulses are equally spaced, they can be interpreted as a single event seen in successive rotations of a neutron star. 
This signature might be accounted for by an object that is dormant or
not generally beamed toward the Earth and whose 
magnetosphere is perturbed sporadically by accretion of material
from an asteroid belt \citep{JimRyan06}.


\subsection{PSR~J1946+24}\label{1946sec}

PSR J1946+24 has $\rm DM = 96$~pc~cm$^{-3}$ and $P = 4.729$~s and was discovered by detecting 4 individual pulses. The intervals between detected pulses in this case were all $ > 20$~s. 
The brightest detected pulse of PSR~J1946+24 has $S/N = 29$ and its dispersion sweep is visible in the time-frequency plane without any stacking (Fig.~\ref{4dynspecs}b). 

\subsection{Summary of Timing Solutions}

Single pulse arrival times collected over multiple follow-up observations were used to obtain partial timing solutions for PSR~J0628+09, PSR~J1909+06, PSR~J1919+17, PSR~J1854+03 and PSR~J1946+24 and verify their estimated periods. The periods of PSR~J0627+16 and PSR~J1909+06 were verified by manually folding raw data and detecting a periodic signal. Finally, despite the fact that PSR~J1928+15 was not detected after the discovery observation, the $\sim 0.403$~s intervals between the three pulses detected from that source differ by only $\sim 2$~ms, which is approximately half the FWHM pulse width and therefore provides a period estimate to that precision.
We are currently timing the six sources which were successfully redetected. Obtaining phase-connected timing solutions will enable us to compare them to other neutron star populations and will yield positions which will facilitate observations at higher wavelengths.


\section{Constraints on Fast Transients }
\label{section_faroutmodel}

The PALFA survey was designed to detect pulsars and pulsar-like objects
by identifying either (or both of) their periodic or single-pulse emission,
as discussed earlier in this paper.  The same data may
also be used to detect 
transient events from other  Galactic and extragalactic sources.
At minimum, the PALFA survey can place limits on the rate and amplitude
distribution of radio transients that are shorter than about 1 second
in duration.  
Possible source classes for short-duration
transient events, to name a few,  include flare stars, magnetar bursts, gamma-ray bursts (GRBs), 
and evaporating black holes.
GRBs have diverse time scales spanning 
$\sim 10$~ms to 1000~s. Mergers of double neutron-star (DNS) binaries  
and neutron star-black hole (NS-BH) binaries are proposed sources for
short bursts, and they may also emit contemporaneous 
radio pulses (\citealt{Paczynski86}, \citealt{Hansen01}).  
Merger rates in the Galaxy
are estimated to be $\sim 1$ to 150~Myr$^{-1}$ for DNS binaries
from pulsar surveys and a factor of 20-30 less for NS-BH binaries from
population synthesis studies (E.g., \citealt{Rantsiou+08}).   Event rates of 
detectable events  depend critically on the luminosity of radio
bursts but are likely to be comparable to the rate of short-duraction
GRBs, and thus  smaller than 
$\sim 1$~event~day$^{-1}$~hemisphere$^{-1}$. 
Another phenomenon that may produce rare, bright pulses 
is the annihilation of primordial black holes (\citealt{Rees77}). 
\cite{Phinney79} estimate the Galactic rate for 
such events to be $< 2$~kpc$^{-3}$~yr$^{-1}$. 
Other possibilities are discussed in \cite{CordesMemo97}. 

Motivated by these considerations, we discuss the constraints that
can be made on transient events from PALFA survey data obtained to date.
First, we note that most --- but not all ---
of the aperiodic events detected 
to date in PALFA data (as reported in \S~\ref{section_objects})
 and in the Parkes Multibeam Survey by \cite{McLaughlin06}, 
appear to be periodic when they are reobserved sufficiently 
to establish a periodicity.   
Additionally, {\em all} events detected so far in both surveys
have dispersion measures that can be accounted for by 
ionized gas in the Galaxy (using the NE2001 model),   
suggesting that the emitting sources are
closely related to the standard Galactic pulsar population. 
By constrast, 
\cite{Lorimer07} recently detected 
a strong (30 Jy), isolated burst with  duration $\sim 5$~ms
that shows DM $\sim 375$~pc~cm$^{-3}$, too large to be accounted for
by modeled foreground material in the Galaxy or by plasma in the
Small Magellanic Cloud, which is somewhat near the direction the 
event was found.


\subsection{Discriminating Celestial Events from RFI}

In addition to detecting radio sources on axis in data streams from
each of the seven beams of the ALFA system, it is possible to detect
strong events off axis with low but non-zero sensitivity over 
almost the entire hemisphere centered on the zenith.  Consequently,
it is possible to detect relatively low-rate but very bright transients
off axis as well as the much weaker but higher rate single pulses 
that we have detected in the on-axis parts of the beam pattern.   

To establish that a given
event is celestial in nature, we need to distinguish it from terrestrial
interference.  For on-axis events, this classification process
 is aided by the directionality
of the seven beams and the expectation that most real events are likely to be
seen in three or fewer  contiguous beams rather than in all seven beams.   
Bright events may violate this expectation, however, because 
near-in sidelobes are fairly large (c.f. Figure 1 of \citealt{palfa1}).
Celestial and terrestrial radio signals can also enter the seven beams of the 
ALFA receiver indirectly --- far off the nominal pointing axis --- through 
reflection and scattering off telescope
support structures, as with any antenna.  
Extremely bright signals (celestial or terrestrial) may be detected 
after scattering into one or more beams, and the effective gain is 
approximately that of an isotropic radiator with small gain  for events
incident far off axis from the pointing direction.
For the Arecibo telescope, which has an intricate support structure with
many possible scattering surfaces, such wide-angle incidence is likely 
to provide multiple paths for radiation to enter the feed optics.  
Radiation arriving along multiple paths will constructively and
destructively interfere
across ALFA's focal plane, so that the event's strength will be nonuniform
in the different data streams.  

When radiation enters the telescope optics from wide angles,
it is difficult to distinguish 
celestial from terrestrial signals based on 
the nominal pointing direction of the telescope and on whether 
they are detected in a single  beam, a subset of beams, or all beams.
Conceivably, extremely bright events from a single radio source could be
emitted at very rare intervals that would  be detected when the telescope is
pointed nominally at directions separated by tens of degrees (but within
the same hemisphere). Such sources and events will be exceedingly 
difficult to disentangle from terrestrial RFI.

Of course the most powerful confirmation
of a particular event is the detection of similar events in reobservations
of the same sky position.  For most of the events reported in 
\S~\ref{section_objects} and
by \cite{McLaughlin06}, redetections have been made in this manner.
They have also relied on establishing that their signatures in the
frequency-time plane conform to what is expected. 
The simplest celestial signals are narrow
pulses that show differential arrival times 
in accord
with the cold-plasma dispersion law  (e.g. \citealt{JimMaura03}). 
Celestial objects
may also show   the effects of multipath propagation
through the interstellar medium (ISM) either in the frequency 
structure that is produced in the spectra  of nearby pulsars or
in the asymmetric broadening of the pulses from  more distant objects.   
For cases where
spatial information associated with the nominal telescope pointing
direction is insufficient to constrain the celestial nature of 
a particular event, we must rely especially  on a detailed study of the 
event's structure in the time-frequency plane.  
Indeed the time-frequency structure of
the event discussed by \cite{Lorimer07} was a key part of the argument
for its classification as a celestial event.

\subsection{Simple Model for ALFA Beam Patterns}

A given reflector and feed antenna have a net
antenna power pattern with a main lobe and near-in sidelobes 
that is similar to 
an Airy function with a main-lobe angular  width  $\sim \lambda/D_a$,
where $D_a$ is the effective reflector diameter.
The wide-angle part of the antenna pattern has 
far-out sidelobes that  are independent of $D_a$ and have average gains 
$G \sim 1$, similar to that of an isotropic antenna. 
However peaks in the far-out sidelobe pattern can have 
gains significantly larger than $G = 1$, corresponding to special directions
where scattering from the telescope support  structure is especially efficient. 
For the Arecibo 
telescope, we expect sidelobes to have the same general properties, with 
the caveat that support structures will introduce further complexity in 
the far-out sidelobe pattern that varies as the telescope is used
to track a source. 
Tracking a sky position is done by rotating the azimuth arm and moving
the Gregorian dome along the azimuth arm, thus changing the 
scattering geometry.
We therefore expect the sidelobe structure to change
significantly with azimuth and elevation of the targeted position on 
the sky. For the seven-beam ALFA system, the main beams by design
sample different parts of the sky, as do the near-in sidelobes. 
At wider angles, the power pattern of each feed overlaps the others
but with considerable variation in gain from feed to feed.

To derive simple constraints on burst amplitudes and rates, we expand
the gain for each feed 
into inner beam and wide-angle terms:
\be
G\left(\theta,\phi \right) = G_{\rm max} P_n\left(\theta,\phi \right) + \left(1 - \eta_B\right),
\ee
where $\theta$ and $\phi$ are polar and azimuthal angles, respectively; 
$P_n(\theta,\phi)$ is the power pattern for the main beam and 
near-in sidelobes; $\eta_B$ is the main-beam efficiency,
 and $(1-\eta_B)$ is the average level of the 
far-out sidelobes. The boresight gain is
\be
G_{\rm max} = \frac{4\pi A_e}{\lambda^2} = \frac{4\pi}{\Omega_A} = \frac{4\pi\eta_B}{\Omega_{\rm MB}},\label{eqn_Gmax}
\ee
where $ A_e$ is the effective telescope area, and $\Omega_A$ and  $\Omega_{\rm MB} = \eta_B \Omega_A$ are the solid
angles of the antenna power pattern and of the main beam, respectively.  
The system equivalent flux density is
\be
S_{\rm sys} = \frac{T_{\rm sys}2k}{A_e} = \frac{8\pi k T_{\rm sys}}{\lambda^2 G_{\rm max}},\label{eqn_Ssys},
\ee
where $k$ is the Boltzmann constant and $T_{\rm sys}$ is the system temperature. The minimum detectable flux density is 
\be
S_{\rm min} = \frac{m S_{\rm sys}}{\sqrt{N_{\rm pol} \Delta f W}},
\label{eqn_Smin}
\ee
which is a function of the event duration, $W$, as well as system parameters. 
It is useful to define 
$S_{\rm min,1}$ as the minimum detectable flux density for unit gain 
using Eq.~\ref{eqn_Ssys} with $G_{\rm max} = 1$. Substituting into
Eq.~\ref{eqn_Smin} yields,  for two polarizations and $\Delta f = 100$~MHz, 
\be
	S_{\rm min, 1} = 10^{4.9}~{\rm Jy}~
	\left(\frac{m}{5} \right)
	\left(\frac{10~{\rm ms}}{W} \right)^{1/2}
	\left(\frac{T_{\rm sys}}{30~{\rm K}} \right);
\ee
we note that for some directions, the system temperature can differ
from the fiducial 30~K used.  

\section{Constraints on Event Rates and Amplitudes}
\label{section_constraints}

We now derive limits that can be placed on the rate
and amplitudes of transient events
using our simple
model for the antenna power pattern.  
Let $\zeta$ be the event rate per unit solid angle from a population
of sources, which implies a total number 
$\zeta T_{\rm obs} \Delta\Omega$  
in a total observation time $T_{\rm obs}$.
If no events are detected, 
we can place an upper limit on the rate 
$\zeta_{\rm s, max}$ for minimum flux density, $S_{\rm min}$. 
We treat the response of the antenna power pattern
as  azimuthually symmetric about the bore axis, 
so $G(\theta, \phi) \to G(\theta)$, and calculate
the gain as a function of polar angle from the bore axis. 
For each annulus of solid angle $\Delta\Omega$, the upper bound
on the rate is
\be
\zeta_{\rm s,max} \leq \frac{1}{M\Delta\Omega T_{\rm obs}},
\ee
for flux densities greater than
\be
S_{\rm min} = \frac{S_{\rm min,1}}{ G(\theta)}.
\ee
The multiplier $M$  accounts for whether 
a given ALFA feed and receiver combination samples the same patch of
sky or not.  For the main lobes, $M=7$ while for the far-out sidelobes,
$M=1$.  

We evaluate fiducial  values by integrating respectively
over the main beams and over the remainder of the power patterns.
For a main-lobe
beam width of 3.5 arc min,
$\Omega_{\rm MB} \approx 10^{-2.6}$~deg$^2$
and $G_{\rm max} \approx 10^{7.0}$. Using  the 
average gain over the main beam out to the half-power point,
$\langle G\rangle \approx G_{\rm max}/(2\ln 2)$, we have
\be
S_{\rm min} = \frac{2\ln 2 S_{\rm min,1}}{G_{\rm max}}
	\approx 11~{\rm mJy}~ 
	\left(\frac{m}{5} \right)
	\left(\frac{10~{\rm ms}}{W} \right)^{1/2}
	\left(\frac{T_{\rm sys}}{30~{\rm K}} \right)
\ee
and 
\be
\zeta_{\rm s,max} = \frac{1}{7\Omega_{\rm MB}T_{\rm obs}}
	\approx 
	0.12~{\rm hr^{-1}~deg^{-2}} 
	\left(\frac{461~{\rm hr}}{T_{\rm obs}}\right).
\ee
Integrating over wide angles, and using
$\eta_{\rm B} \approx 0.7$, we estimate 
\be
S_{\rm min} \approx \frac{S_{\rm min,1}}{1 - \eta_B}
	\approx 10^{5.4}~{\rm Jy}~  
	\left(\frac{m}{5} \right)
	\left(\frac{10~{\rm ms}}{W} \right)^{1/2}
	\left(\frac{T_{\rm sys}}{30~{\rm K}} \right),
\ee
where the far-out sidelobes have an average gain of $1 - \eta_B$, 
and
\be
\zeta_{\rm s,max} = \frac{1}{2\pi\epsilon T_{\rm obs}}
	\approx  \frac{10^{-7.0}}{\epsilon}~{\rm hr^{-1}~deg^{-2}}
        \left(\frac{461~{\rm hr}}{T_{\rm obs}}\right).
\ee
where $\epsilon \approx 1$ is the fraction of the 
hemisphere actually covered by the far-out sidelobes.

\begin{figure}
\centering
\includegraphics[scale=0.4]{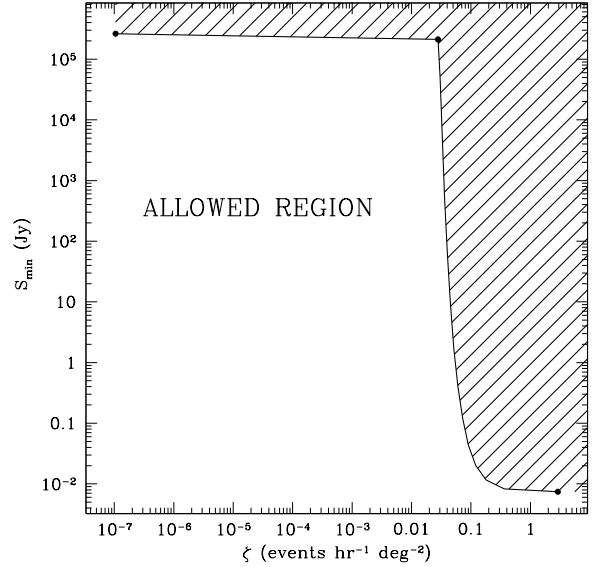}
\caption{Constraints on the single pulse event and minimum detectable
flux density.  
The curve uses a total observation time 
$T = 461$~h and an assumed pulse width of
$W = 10$~ms. 
The three points, from left to right,  correspond to the far-out sidelobes,
the break between the main beam and far-out sidelobes 
(at $\zeta_s \approx 0.03$~h$^{-1}$~deg$^{-2}$),
and the boresight gain.  The shaded region is excluded by the PALFA data.
\label{farout_palfa}
}
\end{figure}

Results for PALFA data are shown in Fig.~\ref{farout_palfa}. 
Within the half-power beam width the constraints are strong on the flux 
density but weak on the event rate while the opposite is true for 
the far-out sidelobes, represented by the part of the curve below 
$\zeta \approx 0.03$~h$^{-1}$~deg$^{-2}$. 
The asympotic value of the curve 
corresponding to the maximum, boresight  gain differs
from the fiducial value calculated above because the latter is an average
over the main beam. 
In our simple model the near-in sidelobes are 
not included explicitly so our curves overestimate $S_{\rm min}$ at
the corresponding polar angles.  However, the limits change rapidly,
so this does not change the salient features of the figure, 
which are the seven order-of-magnitude difference in constrained amplitudes
and the eight order of magnitudes on the event rates between the
main lobs and the far-out sidelobes.  

\subsection{Comparison with Other Results}

A brief summary of our results is that (a) a small number of events 
has been detected from sources that are consistent with their membership 
in the radio pulsar population, but with a much greater deal of
modulation in some cases; (b) no very strong pulses have been
identified in the far-out sidelobes of the Arecibo telescope
and (c) no bursts have been detected that are
extragalactic in origin.   

Here we compare our results with those of 
\cite{Lorimer07}, who surveyed high-Galactic-latitude regions
that included the Magellanic Clouds and who discovered a single reported pulse, with amplitude 30~Jy.
They surveyed $\sim 9$~deg$^2$ with a detection
threshold $\sim 300$~mJy, about 100 times fainter than the detected
pulse amplitude.     
To compare any two surveys, we initially
assume a population of radio sources that
is homogeneously distributed in Euclidean space and we ignore propagation
effects that might limit the detectability of bursts from distant sources. 
The number of events detected $N\propto \Omega_i T D_{\rm max}^3$ for a
survey of duration $T$ that samples instantaneously a solid
angle $\Omega_i$.  
For multibeam systems with $N_{\rm pix}$ pixels, 
$\Omega_i = N_{\rm pix}\Omega_1$, where $\Omega_1$ is the solid angle
of a single beam.  
We assume events have the same peak luminosity and
pulse width $W$ and are detectable to a distance $D_{\rm max}$. 
The ratio of number of detected events in two surveys with total observation times $T_{\rm obs,a}$ and $T_{\rm obs,b}$, and bandwidths $\Delta f_a$ and $\Delta f_b$ is
\ba
\frac{N_b}{N_a} &=& 
\left( \frac{T_{\rm obs,b}}{T_{\rm obs,a}} \right)
\left(	
	\frac{N_{\rm pix,b}\Omega_{\rm 1,b}}{N_{\rm pix,a}\Omega_{\rm 1,a}}
\right)
\left(
	\frac{D_{\rm max b}}{D_{\rm max a}}
\right)^{3}
\nonumber \\
&=&
\left( \frac{T_{\rm obs,b}}{T_{\rm obs,a}} \right)
\left(	
	\frac{N_{\rm b}\Omega_{\rm 1,b}}{N_{\rm pix,a}\Omega_{\rm 1,a}}
\right)
\left(\frac{m_a S_{\rm sys,a}}{m_b S_{\rm sys,b}}\right)^{3/2}
\left(\frac{\Delta f_b}{\Delta f_a}\right)^{3/4}.
\label{eqn_NbNa}
\ea

As pointed out by \cite{Lorimer07}, the same survey that detected
the 30-Jy pulse should have yielded many more detections if the assumptions of
a volume limited sample apply.  Using Eq.~\ref{eqn_NbNa}, the lone
30~Jy pulse implies there should have been 
$N_b / N_a \approx (m_a/m_b)^{3/2} = 10^3$ 
additional detections above a
100-times fainter threshold.  
In the 90~h of follow up observations,
there should have been  
$N_b / N_a \approx (T_{\rm obs,b}/T_{\rm obs,a})(m_a/m_b)^{3/2} \approx 80$
additional detections.    
Comparing the Parkes and PALFA surveys\footnote{For PALFA, we use
7 pixels each of size 3.5~arcmin, 461~h of observations, a threshold 
of $m=5$, a system-equivalent flux density of 4~Jy, and 100 MHz bandwidth.
For the Parkes survey we use 13 beams of 14~arcmin diameter, 480~h,
a threshold of 600, $S_{\rm sys} = 40$~Jy, and 288~MHz bandwidth.}, we estimate that the PALFA survey should have identified $\sim 600$ pulses above threshold. 

The PALFA survey's
null result on extragalactic events is therefore in accord with
Lorimer et al.'s non-detection of events weaker than the single strong
event they reported.   With the assumptions made, these results suggest that
the non-detections of pulses weaker than the 30-Jy pulse are 
inconsistent with the source population having a distribution that is
homogeneous and isotropic.  

\subsection{Caveats}

There are caveats on these results related to the assumptions
about the source population and to the possible role of propagation effects
in limiting detectability.    
The minimum detectable flux density for the PALFA survey
 is 30~mJy for a 5-ms pulse, implying that a pulse of 30~Jy strength
emitted at a distance $D_{\rm event}$ is detectable to  
a distance $D_{\rm max}$ given by 
$D_{\rm max} / D_{\rm event} \approx (30~{\rm Jy} / 0.03~{\rm Jy})^{1/2}
	\approx 32$.
If the lone event detected by Lorimer et al. is 
from a source at the attributed distance
$D_{\rm event} \sim 0.5$~Gpc 
(based on assigning about half of the DM to the
intergalactic medium), the PALFA survey would detect fewer pulses than
estimated above (for constant luminosity) because 
the universe is not old enough and from redshifting of the spectrum
if it is steep in frequency. 
  Nonetheless, a cosmological population
of constant-luminosity sources would extend to $\sim 8$ times the 
attributed distance, implying from the first form of Eq.~\ref{eqn_NbNa}  
that $\sim 7$ pulses should have been detected in the PALFA survey, on average, while $\sim 500$ weaker pulses should
have been seen in the discovery survey with Parkes.  
The difference in
yield under this alternative scaling results from the fact
that the PALFA survey can detect sources well
beyond the cosmological population while the Parkes survey is comparatively
shallower (by a factor of ten in flux density, i.e. 300 compared to 30~mJy), 
but covers a factor of 30 greater solid angle.       
Consideration of a broad luminosity function does not alter these 
conclusions qualitatively.  However, we would generally expect the
pulse amplitude distribution to extend to {\em lower} flux densities
(from lower luminosities) given that the sole reported event was much
larger than the threshold and that, therefore, many more pulses are
expected compared to the constant-luminosity assumption.  

Pulse broadening from multipath propagation
in either the ISM or the intergalactic medium (IGM) can also limit the numbers
of events detected in a survey because the broadening increases
with source distance.  The maximum detectable distance
is therefore diminished, as demonstrated for the ISM 
in Figure~\ref{fig:dmax_vs_lp}.  Lorimer et al. argued that the 
strong frequency dependence of the width of the 30-Jy pulse was consistent
with multipath propagation through a turbulent, ionized IGM.  If so, sources
from a greater distance would be less-easily detectable.  In the simplest
scattering geometries, scattering conserves the area of the pulse,
so matched filtering yields a test statistic with 
S/N $\propto (W / \tau_d)^{1/2}$ when the pulse broadening time $\tau_d$ 
is much larger than the intrinsic pulse width (\citealt{JimMaura03}). 
The 30-Jy pulse therefore could have been broadened by an additional
 factor of $10^4$ and the S/N still would have been above threshold. 
The much longer pulses ($\sim 10^4 \times 5$~ms = 50~s) would have been
difficult if not impossible to identify in the time series 
owing to high-pass filtering used to mitigate baseline fluctuations. 
The high-pass filtering was explicit in the Parkes Multibeam data
acquisition but is part of the post-processing in the PALFA analysis. 
Such very broadened pulses could be detected by imaging surveys with relatively short integration times like the NRAO VLA Sky Survey (NVSS) and future surveys with the Australian Square Kilometer Array Pathfinder\footnote{\texttt{http://www.atnf.csiro.au/projects/askap}} or the Allen Telescope Array\footnote{\texttt{http://ral.berkeley.edu/ata}}.  
  
In between this extreme case of 50-s broadening
 and the reported pulse width of $\sim 5$~ms 
is a substantial search volume in which more distant sources could have 
produced detectable events.  Broadening from sources at redshifts greater
than one is increased by the larger electron density but is lessened by
the frequency redshift, so the net scaling of the pulse broadening
with redshift depends on how the scattering material is distributed
along the line of sight.

\section{Conclusions}\label{section_conclusions}

We have examined the current state of knowledge about various classes 
of intermittent radio-loud neutron stars, summarized selection effects 
that may affect the classification of intermittent sources, and 
presented our data processing methods and results from single 
pulse searches performed on PALFA survey data. 
One of our single-pulse discoveries is a relatively persistent emitter that, like many other pulsars, shows a broad amplitude distribution, from which a few bright pulses were detected.
Two sources were detected 
as normal periodic emitters when observed at a lower frequency 
(PSR~J0627+16) and with the more sensitive central beam of the ALFA 
receiver (PSR~J1919+17). Four objects were most likely not detected via 
periodicity search because of their long periods ($P > 2$~s) compared 
to the PALFA observation time of $134-268$~s, and one intermittent 
object (PSR~J1928+15) was discovered by detecting three pulses emitted 
on successive rotations but it was not detected again despite multiple 
reobservations.

Most of the PALFA sources that were discovered solely through
the single-pulse analysis have subsequently been found to be {\it sporadic}
but with measureable and consistent rates of detectable pulses. 
The Parkes survey had similar results.  
However, two sources (the Parkes source PSR J1839$-$01 and the PALFA source
PSR J1928+15) are much more irregular even though an underlying periodicity
has been found in each case.  We therefore distinguish between 
cases where the apparent sporadic behavior is due to a detection threshold
that selects only the strongest pulses from a broad distribution
(e.g. \citealt{Weltevrede06}) and those where emission 
is truly {\it intermittent}.

A major issue is related to the question, if we see transients from a given sky direction only once, how can we figure out what 
phenomena produce them. We must archive radio survey data 
and characterize the wide variety of detected radio transients 
outside the framework of rotating neutron stars if necessary. 
Archived data may yield new discoveries if reprocessed with improved 
search algorithms and a database of transient signals would be an 
invaluable reference as theoretical understanding of energetic radio 
events and prediction of their signatures improves. The PALFA survey 
is taking steps in both directions: raw data are stored in a tape 
archive and data processing products are indexed in a database at 
the Cornell Center for Advanced Computing and currently work is 
under way to make them publicly available via a web 
portal.\footnote{\texttt{http://arecibo.tc.cornell.edu}}

\vspace{0.5cm}
We thank the staff at NAIC and ATNF for developing the 
ALFA receiver and the associated backend systems. In particular, 
we thank Jeff Hagen, Bill Sisk and Steve Torchinsky at NAIC and 
Graham Carrad at the ATNF. We are also grateful to the 
Parkes Multibeam survey team for providing the results incorporated 
in Fig.~\ref{r_vs_np}. This work was supported by the NSF through a 
cooperative agreement with Cornell University to operate the 
Arecibo Observatory. NSF also supported this research through grants 
AST-02-05853 and AST-05-07376 (Columbia University), AST-02-06035 and 
AST-05-07747 (Cornell University) and AST-06-47820 
(Bryn Mawr College). M.A.M., D.R.L. and P.C.C.F. are supported by a 
WVEPSCoR Research Challenge Grant. M.A.M. is an 
Alfred P. Sloan Research Fellow. F.~Crawford is supported by grants 
from Research Corporation and the Mount Cuba Astronomical Foundation. 
J.W.T.H. is a NWO Veni Fellow. 
Pulsar research at UBC is supported by NSERC and the Canada 
Foundation for Innovation. I.H.S. acknowledges support from the ATNF 
Distinguished Visitor program and the Swinburne University of Technology Visiting Distinguished Researcher Scheme. 
L.E.K. held an NSERC Canada 
Graduate Scholarship while most of this work was performed. 
V.M.K. holds a Canada Research Chair and the Lorne Trottier Chair, 
and is supported by NSERC, FQRNT, CIFAR, and by the 
Canada Foundation for Innovation. The Parkes telescope is part of 
the Australia Telescope, which is funded by the 
Commonwealth Government for operation as a National Facility 
managed by CSIRO. NRAO is a facility of the NSF operated under 
cooperative agreement by Associated Universities, Inc. 
Research in radio astronomy at the NRL is supported by the 
Office of Naval Research.

\end{document}